\documentclass[prl,reprint,showpacs,twocolumn,superscriptaddress]{revtex4-1}
\usepackage{amsmath}
\usepackage[dvips]{graphicx}
\usepackage{bm}
\usepackage{epsfig}
\usepackage{natbib}
\usepackage{upgreek}
\usepackage{pstricks}

\newcommand{\tref}[1]{Table~\ref{#1}}

\begin{document}

\title{Ion neutralisation mass-spectrometry route to radium monofluoride (RaF)}
\author{T.A.\ Isaev}

\affiliation{Clemens-Sch{\"o}pf Institute, 
             TU Darmstadt, Petersenstr. 22, 64287 Darmstadt, Germany}

\author{S.\ Hoekstra}

\author{L.\ Willmann}
\affiliation{Kernfysisch Versneller Instituut, University of Groningen, 
             Zernikelaan 25, 9747 AA Groningen, The Netherlands}      
              
\author{R.\ Berger}
\email{robert.berger@tu-darmstadt.de}
\affiliation{Clemens-Sch{\"o}pf Institute, 
             TU Darmstadt, Petersenstr. 22, 64287 Darmstadt, Germany}

\date{\today}
\pacs{31.15.bw, 31.30.-i, 37.10.Mn, 34.50.Fa, 12.15.Mm, 21.10.Re}

\begin{abstract}
The diatomic molecule radium monofluoride (RaF) has recently been proposed
as a versatile probe for physics beyond the current standard model.
Herein, a route towards production of a RaF molecular beam via radium ions
is proposed. It takes advantage of the special electronic structure
expected for group 2 halides and group 2 hydrides: The electronic ground
state of neutral RaF and its monocation differ in occupation of a
non-bonding orbital of $\sigma$ symmetry. This implies similar equilibrium
distances and harmonic vibrational wavenumbers in the two charge states and
thus favourable Franck--Condon factors for neutralisation without
dissociation in neutralising collisions. According to the calculated
ionisation energy of RaF, charge exchange collisions of RaF$^+$ with sodium
atoms are almost iso-enthalpic, resulting in large cross-sections for the
production of neutral radium monofluoride.
\end{abstract}

\maketitle

We have shown recently that radium monofluoride (RaF) can be considered a
promising open-shell molecular candidate which is well-suited for a wealth
of tests on physics beyond the standard model \cite{Isaev:10a,Isaev:2013}.
This is a result of the following properties of the RaF system. The
molecules is proposed to be amenable for direct cooling with lasers
\cite{Isaev:10a} and it shows favourable relativistic enhancement factors
\cite{Isaev:2013}, a particular open shell structure of diatomic
molecules with a ${}^{2}\Sigma_{1/2}$ ground state
\cite{labzowsky:1978,sushkov:1978,kozlov:1985,Isaev:10a}, and nuclear
octapole deformations for certain radium isotopes
\cite{Auerbach:96,Spevak:97,gaffney:2013}. No experimentally obtained
information on molecular parameters of diatomic RaF in any of its charge
states appears to be available in the open literature and thus quantum
chemically computed data are the only source of information. 

Herein we propose a route to gain first access to information on neutral
radium monofluoride via intermediate production of RaF$^+$, which is
subsequently neutralised by charge exchange in collision with a suitable
chosen collision gas or by interaction with surfaces that provide the
adequate work function for an iso-enthalpic electron transfer. As the
degree of electron-transfer-induced dissociation should be kept at a
minimum, favourable Franck--Condon factors (FCFs) for 
neutralisation are required. This is related to one of the general
prerequisites for molecules to serve as potential candidates for being
cooled with lasers, which has recently been successfully demonstrated for
the molecule SrF \cite{Shuman:10}. According to our classification scheme
of molecules that should be amenable to being cooled directly with lasers
\cite{Isaev:10a}, changes in occupation of non-bonding orbitals disturb the
molecular structures of diatomic molecules only to a small degree, such
that favourable FCFs can be expected in these transitions. As similar
considerations hold for the efficiency of neutralisation-reionisation and
charge-reversal transitions in molecular systems, which have found
a wealth of applications \cite{wesdemiotis:1987,trelouw:1987},
the same classification can also be applied to identify particular suitable
compounds for charge-state changing transitions, which can be light-induced
or collision-induced. 

As the lowest energy ionisation step for radium monofluoride can be
expected to be due to the removal of an electron in a non-bonding orbital
of this heavy elemental compound (case 1 and 3 in Ref.~\cite{Isaev:10a}),
the corresponding closed-shell RaF$^+$ in its ground state is expected to
feature quite similar bond length and vibrational frequencies. In this
brief note, we demonstrate with explicit four-component Fock-Space
Coupled-Cluster (4c-FSCC) calculations using the Dirac program package
\cite{DIRAC:11} that this indeed true. Details of these computations are
described in the supplementary material.

RaF$^+$ can in principle be formed in reactive collisions of radium ions
with a suitable fluorine containing compound. This expectation is fuelled by
the work of Armentrout and Beauchamp \cite{armentrout:1981} who generated
UF$^+$ for instance by reaction of U$^+$ with methyl fluoride (CH$_3$F) to
yield the desired molecular cation in an exothermic process, such that
cross sections decreased with increasing collision energy.  Alternatively,
reaction of uranium ions with tetrafluoro-silane (SiF$_4$) afforded UF$^+$
in an endothermic process, such that cross sections increased with
increasing collision energy. These and other fluorinating agents
such as SF$_6$, PF$_5$ and NF$_3$ should yield also the radium monofluoride
monocation, with the choice of reagent offering tunability in this respect.
According to the computed dissociation energies (see \tref{raf}),
SiF$_4$ should provide access to RaF$^+$ in an endothermic
process, whereas the analogous reaction with CF$_4$ is expected to proceed
almost thermoneutral.

%
%

To assess the viability for an undissociated neutralisation of RaF$^+$ we
computed the potential energy curve of electronic ground state RaF$^+$,
compared this curve to the one of neutral RaF in its electronic ground
state and estimated the FCFs for transition to the lowest vibrational states 
of the neutral in the electronic ground state. The predicted molecular 
parameters are given in \tref{raf} along with the data computed 
previously for RaF.

\begin{table}
\caption
 {Estimated molecular spectroscopic parameters for the
  electronic ground state $^1\Sigma_{0}$ of ${}^{226}$RaF$^+$ from
  FS-CCSD calculations (see supplementary material for details) and comparison 
  to previously calculated parameters of neutral RaF.}
 \label{raf}
\begin{tabular}{llllll}
& $R_\mathrm{e}/a_0$ & $\tilde{\omega}_\mathrm{e}/\mathrm{cm}^{-1}$  &
$\tilde{D}_\mathrm{e}/(10^4 \mathrm{cm}^{-1})$     &
$\tilde{T}_\mathrm{e}/(10^4 \mathrm{cm}^{-1})$ & \\\hline
$^1\Sigma_{0}$   & 4.14 & 502 & 4.51     & 4.00  & this work       \\
$^2\Pi_{1/2}$    & 4.29 & 428 & 4.24$^a$ & 1.33  & Ref.~\cite{Isaev:2013}\\
$^2\Pi_{1/2}$    & 4.24 & 432 & 3.13     & 1.40  & Ref.~\cite{Isaev:10a} \\
$^2\Sigma_{1/2}$ & 4.29 & 431 & 4.26$^a$ & \phantom{0.0}0  & Ref.~\cite{Isaev:2013} \\
$^2\Sigma_{1/2}$ & 4.24 & 428 & 3.21     & \phantom{0.0}0  & Ref.~\cite{Isaev:10a} \\\hline
\multicolumn{6}{l}{\footnotesize a) this work} \\
\end{tabular}
\end{table}

\begin{figure}[h]
\begin{pspicture}[](1.5,0)(8.,7.5)

    \rput[bl](0.,0.5){ \includegraphics[width=\linewidth]{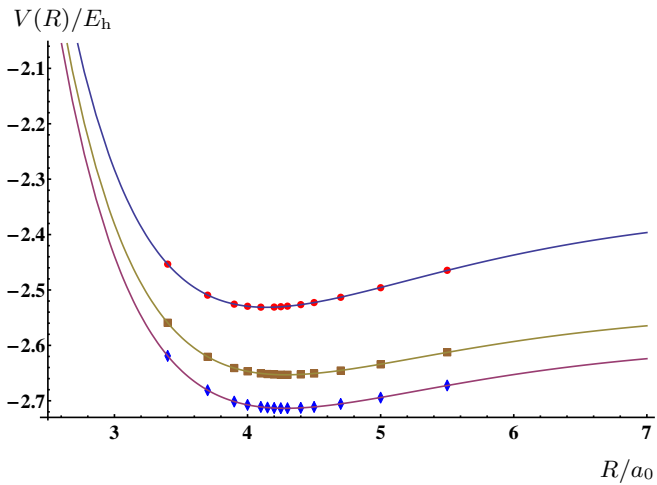}}

    \rput[bl]{0}(0.2,6.){$V(R)/ E_\mathrm{h}$}
    \rput[bl]{0}(8,0.0){$R/ a_\mathrm{0}$}

\end{pspicture}
\caption{(Color online) Morse potential fit of the calculated 
potential energy points for the electronic ground state $^1\Sigma_{0}$
of RaF$^+$ (upper curve, circles) as well as the electronic ground state 
$^2\Sigma_{1/2}$ (lowest curve, diamonds) and electronically first excited
state $^2\Pi_{1/2}$ (squares) of RaF.}
\label{plot}
\end{figure}

According to the data presented in \tref{raf}, favourable FCFs exist for the
neutralisation step of RaF monocation (anharmonic Morse oscillator
estimates for the FCFs are $0.49$ ($0 \rightarrow 0$); $0.34$ ($0
\rightarrow 1$); $0.12$ ($0\rightarrow 2$) and imply a nearly quantitative
FC transition to bound levels of the neutral, even for vibrational
temperatures of 3000~K in RaF$^+$).  Charge-exchange cells with selected
alkali vapours are routinely used to neutralise low-energy ions beams with
up to unit efficiency \cite{klose:2012}.  As the electron affinity of
RaF$^+$, and conversely the ionisation energy of neutral RaF, are computed
to be about 5~eV, collision with sodium vapour ($I = 5.14~\mathrm{eV}$) can
facilitate almost thermoneutral charge exchange between RaF$^+$ and sodium
with comparatively large cross sections for yielding RaF in its electronic
ground state.  Interestingly, the lowest excited doublet state of RaF has
an almost identical potential energy curve as the electronic ground state
and thus similarly favourable FCFs for the neutralisation process.  As
atomic caesium features an ionisation energy of about $I=3.89~\mathrm{eV}$,
a collision with a vapour of caesium can provide direct access to RaF in the
$^{2}\Pi_{1/2}$ excited state in a slightly endothermic reaction. The
corresponding fluorescence to the electronic ground state might then be
monitored to give first indications of the properties of neutral RaF being
produced in this collision. This might also be considered as the first step
towards an optical cooling of RaF with lasers. If a laser beam is aligned
along the molecular beam of neutral RaF, a more detailed investigation
(collinear spectroscopy, see for instance \cite{ahmad:1988,klose:2012}) of
molecular properties can be facilitated.  An indirect detection of RaF can
be achieved by subsequent collisional reionisation of RaF in a
neutralisation-reionisation type mass spectrometry experiment
($^+$NRMS$^+$). This type of experiment could for instance be realised at
CERN's Isotope Separator On Line (ISOLDE) facility: the typical yield of
Ra$^+$ achieved there is about $10^8$ ions/$\upmu\mathrm{C}$
\cite{ISOLDEdata,Stora:priv}.  According to \cite{Stora:priv} a beam of
$^{222}$RaF$^+$ was also observed in July of 2007 with an intensity of
$2\cdot10^6$ ions/$\upmu\mathrm{C}$ while bombarding a uranium carbide
target with high-energy protons.  Taking for an estimate of the RaF$^+$
beam intensity 1/10 of the yield of the corresponding Ra isotope from the
database \cite{ISOLDEdata}, one can expect an intensity of
$^{225,226}$RaF$^+$ of the order 10$^7$~ions/$\upmu\mathrm{C}$ (or
$10^7$~ions/$\mathrm{s}$ for a typical proton beam intensity of
$1~\upmu\mathrm{A}$). 

To summarise the possible route towards a molecular beam of neutral radium
fluoride: Charged radium isotopes formed in spallation experiments are
reacted with a fluorine containing collision or buffer gas to afford
RaF$^+$, which can subsequently be mass selected and neutralised in a
collision with a vapour of sodium or a suitable surface. If a material with
sufficiently low work function exists, generation of RaF in an
electronically excited state is also possible, such that RaF can be
identified by emission of a fluorescence photon, as the initiation of the
first cycle in a cooling procedure with lasers. This scheme should give
access to neutral RaF, which -- after cooling and stopping either with
lasers or by Stark deceleration -- can be subject to a detailed search
for physics beyond the standard model.

\paragraph*{Acknowledgements}
We are indebted to T. Stora for discussions and particularly thankful for
numerous discussions with various participants at the 2010 ECT* workshop on
"Violations of discrete symmetries in atoms and nuclei" in Trento.
Computer time provided by the Center for
Scientific Computing (CSC) Frankfurt is gratefully acknowledged.

%
\end{document}